\documentclass[aps,preprint]{revtex4}

\usepackage{graphicx}

\begin{document}

\date{\today}

\title{Equilibrium states in open quantum systems} 

\author{
Ingrid Rotter\footnote{email: rotter@pks.mpg.de}}
\affiliation{
Max Planck Institute for the Physics of Complex Systems,
D-01187 Dresden, Germany  }

\vspace*{1.5cm}

\begin{abstract}
The aim of the paper is to study the question whether or not
equilibrium states exist in open quantum systems that are embedded in
at least two environments and are described by a non-Hermitian
Hamilton operator $\cal H$. The eigenfunctions of
$\cal H$ contain the influence of exceptional points (EPs) as well as
that of external mixing (EM) of the states via the environment. As a
result, equilibrium states exist (far from EPs). They are different
from those of the corresponding closed system. Their wavefunctions are
orthogonal although the Hamiltonian is non-Hermitian.

\end{abstract}

\maketitle

\section{Introduction}
\label{intr}

In many recent studies, quantum systems are described by a
non-Hermitian Hamilton operator $\cal H$. Mostly, only the eigenvalues
of $\cal H$ are considered because the main interest of these studies
consists in receiving an answer to the question whether the
eigenvalues of the non-Hermitian Hamiltonian are real or complex
\cite{bender}. Also the conditions for the change from real to complex
eigenvalues and vice versa are considered in many papers. In other
studies, calculations for realistic small quantum systems are
performed by using the complex scaling method \cite{mois}.
Newly, also topological insulators are considered in the presence of
non-Hermiticity \cite{lee}. In the description of realistic systems,
the non-Hermiticity of the Hamiltonian arises from the embeddening of
the system into an environment as explained in \cite{top,ropp}. It is
derived from the full Hamiltonian describing  the system together with
the environment which is, of course, Hermitian.

In order to describe the properties of a realistic quantum system one
needs not only the eigenvalues of the Hamiltonian but also its
eigenfunctions. This is nothing but the longtime experience obtained
from standard quantum mechanics calculations performed with Hermitian
operators. There is no reason why this should be different in the
description of the system by means of a non-Hermitian Hamiltonian.

So far, the characteristics of the eigenvalues together with those of
the eigenfunctions of a non-Hermitian Hamiltonian $\cal H$ 
are studied for a small open quantum system 
in the papers of only one group, see e.g. \cite{top}. The theoretical
results are compared with those known experimentally \cite{ropp}. The
theoretical studies are performed in the same manner as those in
standard quantum mechanics with the only difference that the Hermitian
Hamiltonian $H$ is replaced by the non-Hermitian operator $\cal
H$. The results allow us therefore to find a direct answer to the
question whether or not it is really necessary to describe the system
(in a special situation) by a non-Hermitian Hamilton operator. The
results show that a unique answer does not exist. It will depend
rather on the conditions under which the system is considered.
 
Of special interest is the challenging question whether or not an
equilibrium state exists in an open quantum system that is embedded in
at least two environments. The challenge of this question arises from
the fact that it combines two conflicting concepts. On the one hand,
the system is open meaning that its properties are not fixed and will
vary under the influence of the environment. On the other hand, the
system is expected to be in equilibrium. It is therefore no wonder
that this question is not at all considered, up to now.

The present paper is aimed at a study of this question in the
framework of the non-Hermitian formalism. For that purpose, we will
sketch some results obtained by using the formalism worked out in
\cite{top}. We will then consider the parametric evolution of an open
quantum system and the possibility to
form an equilibrium state. In any case, such a state is expected to be
different from the equilibrium state of the corresponding closed
system (described by the Hermitian Hamilton operator $H$).
  
The paper is organized in the following manner. In
Sect. \ref{nh-vers-h}, the non-Hermitian formalism is shown to differ
from that of the Hermitian formalism by  two basic features: (i)  the
existence of singular points which are called mostly exceptional
points (EPs) \cite{kato} and (ii)  the possibility of a mixing of the
eigenfunctions of $\cal H$ via the environment called usually external
mixing (EM) \cite{pra95}. In the following Sect. \ref{ent1}, the
information entropy is defined while in Sect. \ref{evol}, the possible
formation of an equilibrium state in an open quantum system (coupled
to at least two environments) is considered. 
Sect. \ref{concl0} is devoted to the different aspects of the question
whether or not it is really necessary to consider the non-Hermiticity
of the Hamiltonian when describing an open quantum system. Here, we
consider the most common case that the system is coupled to more than
one environment. The conclusions are contained in the last
Sect. \ref{concl}. In appendix \ref{a1}, a few experimental results
are listed which cannot be explained in the framework of Hermitian 
quantum physics.

\section{Non-Hermitian versus Hermitian formalism}
\label{nh-vers-h}

In standard quantum theory, the Hamilton operator of a many-body 
system is assumed to be Hermitian. Its
eigenvalues are real and provide the energies of the states. The
lifetimes of the states cannot be calculated directly. They are obtained
usually from the probability according to which the particles may tunnel 
from inside the system to outside of it. 
The Hermitian Hamilton operator describes a closed system. 

The standard quantum theory 
based on a Hermitian Hamiltonian,
has provided very many numerical results which agree
well with the experimentally observed results. A few experimental
results could however not be explained in spite of much effort. They
remained puzzling (see appendix \ref{a1} for a few examples). 

Full information on a quantum system can be obtained only when it 
is observed either by a special measurement devise or, in a natural
manner, by the environments into which the system is embedded. 
Such a system is open and is described best by a Hamilton operator 
which is non-Hermitian \cite{top}.

Quantum systems are usually localized in space, meaning that they 
have a finite extension with a shape characteristic of them. 
The environments are however infinitely
extended. The environments of the open quantum system
can, generally, be parametrically varied and allow thus a 
parameter dependent study of the properties of quantum systems.
Studies of such a type are performed by now 
theoretically as well as experimentally on very different open
quantum systems \cite{ropp}. They provided very many  interesting
results which are basic for a deeper understanding of quantum physics.

From a mathematical point of view, the non-Hermitian formalism is much
more complicated than a formalism which is based on a Hermitian
Hamilton operator. Most important problem  is the existence of
singularities \cite{kato} that may appear in the non-Hermitian formalism.
At these singular
points, called mostly exceptional points (EPs),
two eigenvalues of the Hamiltonian coalesce and --
even more relevant for the system properties -- 
the eigenfunctions show deviations
from the expected ones not only at these points but in a certain
finite neighborhood around these points \cite{top,pra95}. The
deviations are caused by the nonlinearities involved in the equations
at and near to the EPs.
 
At an EP, the trajectories of the  eigenvalues ${\cal E}_{1,2}
\equiv  E_{1,2} + i\Gamma_{1,2}/2$ of $\cal H$ \cite{comment2} of the
two crossing states do not move linearly as a function of a certain
parameter. The eigenvalue trajectories are rather exchanged: the
trajectory of state 1 continues to move as the trajectory of state 2
and vice versa. The same holds true for the trajectories of the
eigenfunctions.

The motion of the eigenvalue trajectories is influenced by an EP not
only at the position of the EP but also in its vicinity. Here, the
Im(${\cal E}_i$) may be exchanged while the corresponding Re(${\cal
  E}_i$) are not exchanged; or the other way around.
 The first case (exchange of the widths) can be traced up to discrete
 states \cite{ro01} which is nothing but the Landau-Zener effect known
 very well (but not fully explained)  in Hermitian quantum physics.
 The trajectories of the eigenfunctions show a behavior which corresponds 
 to that of the eigenvalue trajectories as illustrated first in \cite{ro01}.
 In any case, a linear motion of the eigenvalue trajectories occurs
 only far from an EP.
 
Another important problem is the fact that all states of the system
may interact via
the environment. This is a second-order process since every state is
coupled to an environment and hence -- via the environment --  
to another state. This process is called usually external mixing 
(EM) of the states \cite{top,pra95}.

A question arising naturally is related to the following problem: 
what are the conditions under 
which the system can be described nevertheless by a Hermitian 
Hamilton operator? In other words:
when it is really necessary to describe the system by 
means of a non-Hermitian Hamilton operator?
 
The results of very many studies on realistic quantum systems are the
following: \\ 
(i) Far from EPs and at low level density (at which every state is
well separated  from neighboring states), the
description of the system by means of a Hermitian Hamilton operator
provides good results. This fact is very well known from 
countless calculations over many years. \\
(ii) Near to EPs, the non-Hermitian formalism provides
results which are counterintuitive. These results  agree (at least
qualitatively)
with experimentally observed puzzling results (see appendix \ref{a1}).\\
(iii) The EM of all states via the environment changes the
eigenfunctions of the
Hamiltonian although EM is a second-order process \cite{top,pra95}.\\
(iv) In approaching an EP, the EM increases up to infinity
\cite{pra95}. It is therefore
impossible for any source (such as light) to interact with the system 
at an EP (for an example see appendix \ref{a1}). 

The answer to the above question is therefore that the non-Hermiticity
of the Hamiltonian has to be taken into account in all
numerical calculations in which the
influence of EPs and EM cannot be neglected. All other cases can be
well described numerically
by means of standard methods with a Hermitian Hamilton operator.

This does however not mean that the Hamiltonian of the
open quantum system is really Hermitian. Quite the contrary: 
the singular EPs cause, in either case, nonlinear processes in a
finite parameter range around their position \cite{pra95}.  These
nonlinear processes being inherent in the non-Hermitian formalism,
determine the (parametric) evolution of an open quantum system.
 
The results obtained from calculations for a
system that is embedded in only one environment,
seem to contradict this statement.
Such a system can be described by taking into consideration the
influence of EPs and EM or by neglecting it. The results are the
same. The reason for this unexpected result is that nonlinear processes
are able to restore the correct results (corresponding to the
calculations with inclusion of EPs and EM) when EPs and EM are
neglected in the calculation. These results feign therefore a good
description of the system by means of a Hermitian Hamilton operator
\cite{pra95}.
It should be underlined however that this occurs {\it only} in the
case when the system is embedded in no more than one well-defined
environment.

New interesting effects occur in systems that are embedded in more
than one environment (what is commonly the case).
A realistic example is the transmission through a small system (e.g. a
quantum dot) which needs, at least, two environments. One of the two
environments is related to the incoming flux and the other one to the
outgoing flux.
Most important new effect is, in this case, the appearance of coherence
\cite{arxiv} additionally to the always existing dissipation in open
quantum systems. Coherence is
correlated with an enhanced transmission. It is therefore of relevance
for applications.

\section{Information entropy and equilibrium states}
\label{ent1}

The properties of open quantum systems depend on many different
conditions simulated mostly by different parameter values.
In many studies, the coupling strength $\omega$
between system and environment
is kept constant at a sufficiently high level. 
By this, the influence of a possible variation of the
value of the coupling strength $\omega$ between system and 
environment on the results is largely suppressed.

According to the results of many calculations,
the main difference between the Hermitian and the non-Hermitian 
formalism of quantum mechanics is
caused by two values: the (mathematical) existence of 
singular EPs \cite{kato} and the (physical) possibility of EM of the
eigenfunctions \cite{pra95}. 
The most challenging question is the following.
Do equilibrium states exist also in a system with EPs and EM?
In other words: do equilibrium states exist
in an open quantum system and how they do look like?

In order to find an answer to this question, we will first define the
information entropy. For this purpose,
let us consider a system consisting of $N$ eigenstates of a
non-Hermitian Hamiltonian $\cal H$, 
the eigenvalues of which are ${\cal E}_i \equiv E_i + i/2~ \Gamma_i$
\cite{comment2}.
Here the  
$E_i$ stand for the energies of the eigenstates and the $\Gamma_i $
for their widths, being inverse proportional to the lifetimes $\tau_i$
of the states. Let us choose $E_i \approx  E_{j\ne i}$ and $\Gamma_i
\ne \Gamma_{j\ne i}$.
Following  Shannon \cite{shannon}, 
the information entropy $ H_{\rm ent}$ for this case can be 
defined  by
\begin{eqnarray}
\label{eq5}
H_{\rm ent} = - \sum_{i=1}^N \; p_i \; {\rm log}_2 \; p_i
\end{eqnarray}
where $p_i$ is the probability to find the state $i$ and 
$ -{\rm log}_2 \; p_i$ is the expectation value 
of the information content of the state $i$ (entropy).
The value $H_{\rm ent}$ is maximum when the different  $p_i$ 
are equal. In this case, the system is in an equilibrium state.

\section{Equilibrium state of an open quantum system}
\label{evol}

An open quantum system (that is, normally, embedded in more than one
environment) can be, if at all,
in an equilibrium state only under two conditions: (i) the system is 
far from an EP and (ii) EM is taken into account.
The first condition arises from the fact that the properties of the
system are not stable when the system is near to an EP. The second
condition corresponds to the fact that EM is inherent in
the eigenfunctions $\Phi_i$ of the non-Hermitian Hamiltonian $\cal H$. 

It should be repeated here that it is, indeed, a challenging question
whether or not an
equilibrium state exists in an open quantum system because 
it combines two conflicting concepts: the system is open (meaning that
its properties are influenced by the environment) and is,
nevertheless, expected to be in equilibrium. It is therefore no wonder
that this question is not at all considered, up to now.
 
To overcome this problem, we introduce the concept of maximum
entropy. It is meaningful also for an open quantum system as will be
shown in the following.
  
Numerical calculations have shown the following (unexpected)
result. Tracing the results for a system embedded in two environments,
by varying a certain parameter, the eigenfunctions
$\Phi_i$ of the non-Hermitian operator $\cal H$ become finally
orthogonal \cite{arxiv}. The system does not evolve further.
  
This result can be understood by taking into consideration that the
evolution of a system described by the non-Hermitian operator $\cal
H$, is driven by nonlinear terms. These nonlinear terms originate from
the EPs and are involved in the corresponding Schr\"odinger
equation. They are responsible for the fact that the system does not
evolve wthout any limit. Instead the evolution of the system occurs, at the
most, up to a certain final state at which the eigenfunctions of $\cal
H$ are orthogonal (instead of biorthogonal).
  
The calculations have shown further that the eigenfunctions $\Phi_i$
of the final state are strongly mixed in the set of wavefunctions
$\Phi_k^0$ of the individual original states,
\begin{eqnarray}
\label{eq6}
\Phi_i ~=~ 
\frac{1}{N}~ \sum_{k=1}^N ~\Phi_k^0 ~\langle \Phi_k^0|\Phi_i\rangle 
\; .
\end{eqnarray}
Here the $ \Phi_k^0$ are the eigenfunctions of the
  non-Hermitian Hamiltonian ${\cal H}_0$ which differs from the full
  operator $\cal H $ by the disappearance of the non-diagonal matrix
  elements. That means the  $ \Phi_k^0$ are the wavefunctions of the
original (unmixed) states while the mixing of the states via the
environment is involved in the $\Phi_i$. 

According to numerical calculations, each of the
 $\Phi_k^0$ appears in (\ref{eq6})  with the same probability in
all the eigenfunctions $\Phi_i$ \cite{arxiv} when the entropy is
maximum.  This means that the final state is an equilibrium state. 
It should be underlined once more that a
general definition of an equilibrium state in an open quantum system
is difficult (or even impossible). It is however always possible 
to find a state with maximum value of entropy.

Summarizing the results obtained for a system that is embedded in at
least two environments, we state the following: (i) the final state of
the evolution of the open quantum system (described by a non-Hermitian
Hamiltonian $\cal H$) is an equilibrium state, and (ii) the
eigenfunctions of $\cal H$ at the final state of the evolution are
orthogonal (and not biorthogonal).

\section{Numerical studies for concrete systems versus 
evolution of open quantum systems} 
\label{concl0}

Let us come back to the question raised in the Introduction: whether
(and when) it is really necessary to describe a realistic open quantum
system by a non-Hermitian Hamilton operator? The results obtained in
different studies are sketched above. They show clearly that the
answer to this question depends on the conditions under which the
system is considered.
 
From the point of view of numerical results  which will 
describe special properties of the system, the answer is the following. 
The system can be described well by a Hermitian Hamiltonian $H$ when
its individual eigenstates are distant from another, so that the
different states of the system do (almost) not influence each
other. This happens at low level density.
 
The system has, however, to be described by a non-Hermitian
Hamiltonian $\cal H$ when the influence of singular points (EPs)
cannot be neglected in the considered parameter range. This occurs
especially at high level density where the different eigenstates are
no longer independent of one another.
 
This means, that the properties of many realistic systems can be
described well by means of a Hermitian Hamiltonian $H$. This statement
corresponds to the longtime experience of standard quantum mechanics.
 
The situation is completely different when general questions such as
the (parametric) evolution of an open quantum system or the formation
of an equilibrium state are considered. In this case, the nonlinear
processes caused by the EPs and involved in the non-Hermitian
formalism are decisive. Among others, they are responsible for the
fact that an open quantum system (embedded in at least two
environments) 
will finally stop to evolve further. The evolution occurs
up to the formation of a certain final state. The wavefunctions of
this final state are orthogonal (and not biorthogonal) although the
Hamiltonian of the system is non-Hermitian. The final state is an
equilibrium state of the whole system (including its second-order
components related to EM). It is different from the equilibrium state
of the corresponding closed system.

\section{Conclusions} 
\label{concl}   
   
The non-Hermitian quantum physics is a fascinating field of research
that not only finds an answer to some long-standing problems of
quantum physics (for examples see  the Appendix and also 
\cite{pra95}). Important is, above
all, that it justifies the use of standard Hermitian quantum physics
for the description of experimental results in very many
cases. Non-Hermitian quantum physics is therefore nothing but a
realistic extension of standard quantum physics.
  
From a mathematical point of view, non-Hermitian quantum physics is
much more complicated than Hermitian quantum physics. An important
difference are the nonlinearities arising from the EPs. They play a
role however only in the vicinity of the EPs.
 
In spite of these differences, the non-Hermitian and the Hermitian
formalism show analogue physical features. An example is the existence
of equilibrium states. As shown in the present paper, equilibrium
states exist not only in closed quantum systems (described by a
Hermitian Hamilton operator) but also in open quantum systems
(described by a non-Hermitian Hamilton operator), though with certain
restrictions. The equilibrium states of an open and a closed system
differ, of course, from one another.

\appendix
\section{Puzzling experimental results}
\label{a1}

Over the years, different calculations with non-Hermitian Hamiltonians
have been performed in order to explain puzzling experimental
results. Here, a few of them will be listed.
 
1. Some years ago, the evolution of the
transmission phase to be monitored across a sequence of 
resonance states has been studied experimentally
in a work where a 
multi-level quantum dot was embedded into one of the arms of an
Aharonov-Bohm interferometer \cite{yacoby}. These
experiments revealed the presence of unexpected regularity in the
measured scattering phases (so-called ``phase lapses"), when the
number of states
occupied by electrons in the dot was sufficiently large.
While this behavior could not
be fully explained within approaches based upon Hermitian quantum
theory, it has  been established that
the phase lapses can be attributed to the non-Hermitian character of
this mesoscopic system, and to changes of the system that occur as the
number of electrons in the dot is varied \cite{laps3}.
The observed regularity arises from the overlap of the many
long-lived states with the short-lived one, all of which are formed in
the regime of overlapping resonance states.

2. An example of an environmentally induced transition that is
different in character to that described above, is the spin swapping
observed in a two-spin system embedded in an environment of
neighboring spins \cite{past1}.
In describing the damped dynamics of the two-spin system, its
interaction with its environment may be taken to be inversely
proportional to some characteristic time ($\tau_{SE}$), which degrades
the spin-swapping oscillations on some ``decoherence time''
($\tau_\phi$).  In the experiment,
 two distinct dynamical regimes were observed. In the 
 first of these, the expected proportionality
 $\omega \propto b$ and $\tau_\phi \propto \tau_{SE}$
 was found as the interaction strength $b$ between the spins
 was increased. This behavior agrees with Fermi's golden rule.  On
 exceeding a critical environmental interaction,
 however, the swapping was instead found to freeze while the
 decoherence rate dropped according to $\tau_\phi^{-1} \propto b^2\,
 \tau_{SE}$. The transition between these two dynamical regimes was not
 smooth, but rather had the characteristics of a critical phenomenon,
 occurring once $\omega$ becomes imaginary. For such conditions,
 damping of the spin motion  decreases with increasing coupling to
 the environment, in marked contrast to the behavior obtained when
 $\omega$ is real. The observed results are related 
 in \cite{past1} to the non-Hermitian Hamiltonian describing the system 
and  to the presence of an EP.

3. The high efficiency of the photosynthesis process (used by plants 
to convert light energy  in reaction centers  
into chemical energy) is not understood in Hermitian quantum physics
\cite{engel}. 
Using the formalism for the description of open quantum systems
by means of a non-Hermitian Hamilton operator, 
fluctuations of the cross section near to singular points 
(EPs) are shown to play the decisive role
\cite{gain}. The fluctuations appear in a natural manner,
without any excitation of  internal degrees of 
freedom of the system. They occur therefore with high efficiency
(and very quickly). The excitation
of  resonance states of the system by means of these fluctuations
(being the second step of the whole
process) takes place much slower than the first one, because 
it involves the excitation of
internal degrees of freedom of the system. 
This two-step process as a whole is highly efficient and the decay
is bi-exponential. 
The characteristic features of the process obtained from analytical
and numerical studies, are the same as those of  light harvesting in
photosynthetic organisms \cite{gain}.

4. Atomic systems can be used to store light and to act therefore as a
quantum memory. According to experimental results, optical storage can
be achieved via stopped light \cite{everett}. Recently, this
interesting phenomenon has been related to non-Hermitian quantum
physics.
It has been revealed that light stops at exceptional points (EPs)
\cite{maily}.
The authors of Ref. \cite{maily} restrict  their study to PT-symmetric
optical waveguides. This restriction is however not necessary. The
phenomenon is rather characteristic of non-Hermitian quantum physics:
the external mixing (EM) of the eigenfunctions of a non-Hermitian
Hamilton operator becomes infinitely large at an exceptional point
(EP) \cite{pra95} such that any interaction with an external source
(such as light) vanishes in approaching an EP.

\end{document}